\documentclass[rmp,pdftex]{revtex4}

{\catcode`\@=11 \global\let\NAT@force@numbers\relax}

\usepackage[pdftex]{graphicx}
\usepackage{bm}
\usepackage{amssymb}
\usepackage{amsmath}

\begin{document}

\title{Fundamental Physics and Relativistic Laboratory Astrophysics \protect\\ with Extreme Power Lasers}
\author{T. Zh. Esirkepov}
\author{S. V. Bulanov}
\address{Kansai Photon Science Institute, Japan Atomic Energy Agency, 8-1-7 Umemi-dai, Kizugawa, Kyoto, 619-0215, Japan}
\begin{abstract}
The prospects of using extreme relativistic laser-matter interactions
for laboratory astrophysics are discussed.
Laser-driven {\it process simulation} of matter dynamics 
at ultra-high energy density is proposed
for the studies of astrophysical compact objects and the early universe.
\end{abstract}
\maketitle

\let\cite\citenum

\section{Introduction}

The advent of chirped pulse amplification (CPA) has lead to a dramatic increase
of lasers' power and intensity (Mourou {\it et al.\/} \cite{Mourou}; Fig. \ref{fig:LaserProgress}). 
Present-day lasers are capable to produce pulses whose
electromagnetic (EM) field intensity at focus is well above 10$^{18}$ W/cm$^2$. 
An electron becomes essentially relativistic in such fields,
so that its kinetic energy is comparable to its rest energy. 
Optics in this regime is referred to as relativistic optics; 
the corresponding EM field is called relativistically strong.
Focused laser intensities as high as $10^{22}$ W/cm$^2$ have been reached in experiments (Bahk {\it et al.\/} \cite{Bahk}).

The increase of laser intensity opened the field of laser-driven
relativistic laboratory astrophysics, where 
laser-matter interactions exhibit relativistic processes 
similar to some astrophysical phenomena. 
Acquiring greater attention from various scientific communities, 
laboratory astrophysics becomes one of the
most important motivations for the construction of the ultra-high power
lasers (Bulanov SV {\it et al.\/} \cite{Bulanov-Frauenworth}).

In 2010 the CPA laser systems operating worldwide reached the total peak
power of 11.5 PW and by the end of 2015 planned CPA facilities, except
exawatt scale projects, will bring the total of 127 PW (Barty \cite{Barty-ICUIL}). The
most powerful laser facilities, {\sf NIF} (LLNL, USA) and {\sf LMJ} (Bordeaux, France),
are intended primarily for inertial thermonuclear fusion. 
The {\sf \cite{NIF}} designed for 1.8 MJ has demonstrated 1.41 MJ with 192 beams of 3$\omega$ 
in August, 2011.
The {\sf \cite{LMJ}} will deliver 1.8 MJ with 240 beams in 2012. 
The {\sf \cite{PETAL}} laser coupled with {\sf LMJ} will produce long pulse beams of 200 kJ
combined with ultra-high intensity beams of 70 kJ for fast ignition
research and fundamental science. Future exawatt scale facilities {\sf \cite{HiPER}}
(High Power laser Energy Research) and {\sf \cite{ELI}} (Extreme Light
Infrastructure) are intended for fundamental science, including high
field science. In the plans of all these facilities there are experiments
for laboratory astrophysics. In the project of {\sf \cite{ELI}}, designed for producing
femtosecond pulses of 70 KJ with the power $>$ 100 PW and the intensity 
$>$ 10$^{25}$ W/cm$^2$, laboratory astrophysics is one of the main motivations.

Laboratory astrophysics
complements and supports astrophysical observations with laboratory
experiments, theoretical and numerical studies, which helps to interpret
observations and allows refining the theory by strengthening its predictive power.
Laboratory astrophysics applications are categorized
into four domains (Savin {\it et al.\/} \cite{LAWP}): 
{\it atomic} and 
{\it molecular} domains, 
{\it dust and ices} domain, 
and {\it plasma} domain. 
They are focused on the interpretation of observed spectra,
identification of astrophysical processes, simulation of some phenomena
experimentally in the laboratory and numerically with computers,
development of instruments and diagnostics for observational astrophysics.
Lasers play important role in all these domains as a tool of diagnostics.
In some cases, lasers drive the processes of interest enabling the
experimental simulation of astrophysical conditions, e.g. radiatively,
magnetically, and kinetically driven laboratory experiments in the plasma domain. 
Recalling recent results from
{\sf \cite{WMAP}} and {\sf \cite{Planck}} spacecrafts,
one can consider the fifth domain, {\it cosmology},
where one of intriguing problems is the state of matter in the early universe. 
Extreme power lasers such as {\sf \cite{ELI}} and {\sf \cite{HiPER}} 
can produce conditions close to the lepton epoch of the early universe.

\section{Laser-matter interaction} \label{sec:key}


\begin{figure}
\includegraphics[scale=0.75]{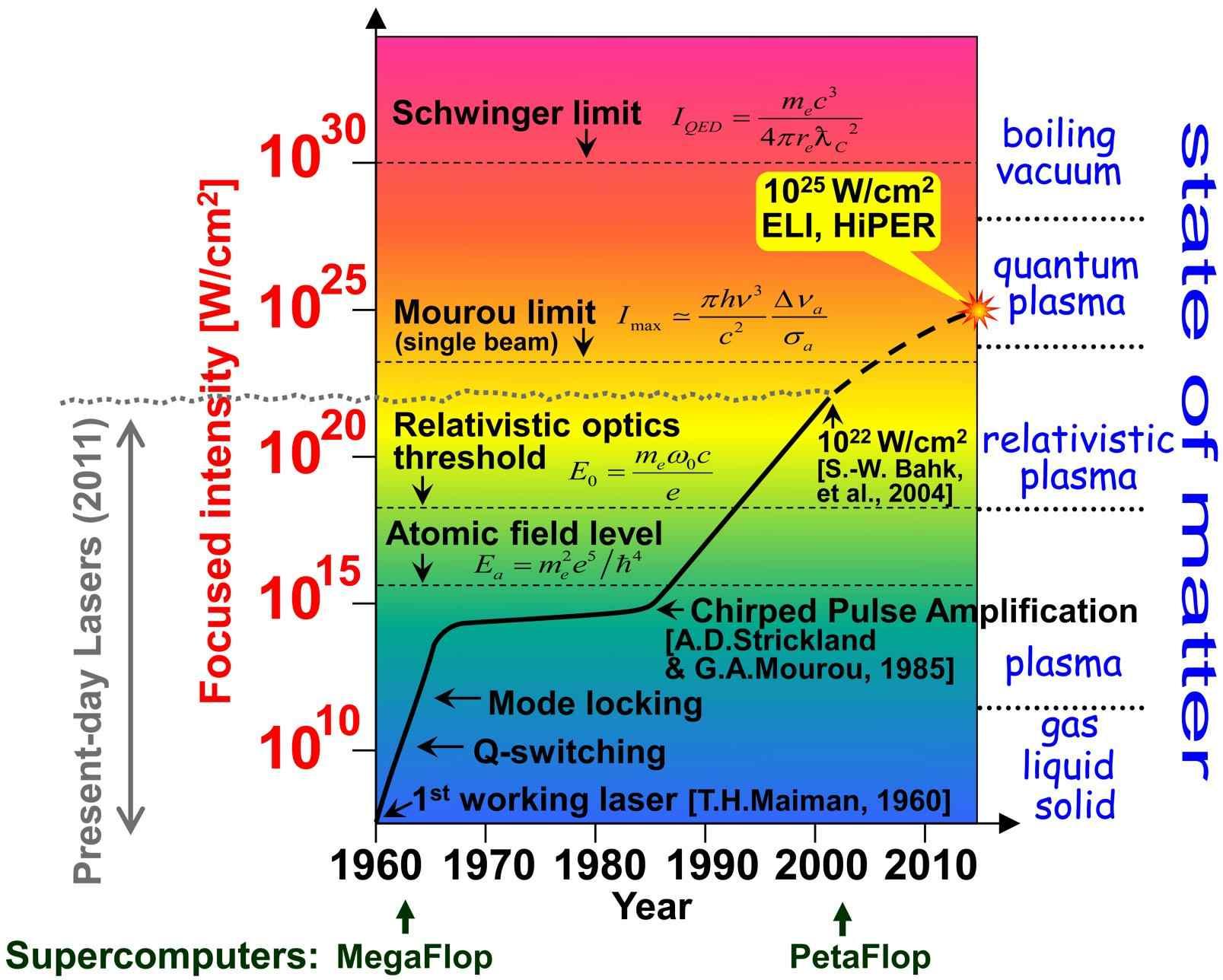}
\caption{\label{fig:LaserProgress}%
The progress of laser technology in terms of focused intensity,
whose increase enables new states of matter. 
The investigation of new regimes of laser-matter interactions
is bolstered up by the supercomputer performance growth.}
\end{figure}

An interaction of an EM wave with matter
is principally determined by its intensity (irradiance),
$I=cE_0^{2}/4\pi$,
where $E_0$ is the electric field amplitude and $c$ is the speed of light in vacuum.
For different thresholds of the laser pulse focused intensity
different nonlinear effects are produced, Fig. \ref{fig:LaserProgress}.
(i) At the intensity of 
$I \simeq 10^{12} - 10^{13}$ W/cm$^2$
two-photon ionization, above-threshold ionization, and double ionization 
have been observed using high-order harmonics (corresponding to wavelengths from 30 to 110 nm)
generated in gases (Sekikawa {\it et al.\/} \cite{bib:HHG}).
(ii) At the intensity of
$I_a = 3.5\times 10^{16}$ W/cm$^2$, 
which is the atomic unit of intensity corresponding to the atomic unit of electric field 
$E_a = m_e^2e^5/\hbar^4$,
the interaction of EM pulses with
atomic gas is highly nonlinear and is accompanied by harmonic generation
(Corkum \cite{Corkum}; Krausz \& Ivanov \cite{Krausz}).
Nonlinearities can start at lower intensities for atoms in the excited states.
Here $m_e$ and $e$ are the mass and charge of electron, and 
$\hbar = h/2\pi$ is the reduced Planck constant. 
(iii) The EM wave with wavelength $\lambda$ and frequency $\omega$
causes the electron dynamics to be relativistic at the intensity of
$I_{rel} = 1.37\times 10^{18}$ W/cm$^2 \times(\mu{\rm m}/\lambda)^2$,
corresponding to the unit value of the dimensionless amplitude of
the wave $a_0 = eE_0/m_e \omega c$ (Mourou {\it et al.\/} \cite{Mourou}).
(iv) At the intensity of
$I_{rad} = 2.6 \times 10^{23}$ W/cm$^2$ $(\mu{\rm m}/\lambda)^{4/3}$, 
the electron dynamics becomes dominated by the radiation reaction 
(Bulanov SV {\it et al.\/} \cite{bib:RRD}).
The corresponding dimensionless amplitude is $a = (3\lambda/4\pi r_e)^{1/3}$, 
where $r_e=e^2/m_e c^2$ is the classical electron radius. 
(v) Quantum effects come into play at the intensity of
$I_{qua} = 6\times 10^{24}$ W/cm$^2$ (Bulanov SV {\it et al.\/} \cite{bib:RRD}).
The corresponding electric field is
$E_{qua} = 4\pi e/3\lambda_C$ , where $\lambda_C = h/m_e c = 2.43\times 10^{-10}$ cm 
is the Compton wavelength of the electron. 
(vi) At the intensity of
$I_S = 2.3\times 10^{29}$ W/cm$^2$, 
corresponding to the critical electric field of
quantum electrodynamics (QED) or so-called Schwinger field
$E_S = m_e^2c^3/e\hbar$,
electron-positron ($e^-e^+$) pairs are created from vacuum 
(Heisenberg \& Euler \cite{bib:QED}; Schwinger \cite{bib:Schw}) 
and Unruh radiation (Unruh \cite{bib:Unruh}) of electron in the electric field
can become dominant (Chen \& Tajima \cite{bib:UnEm}).
Interactions of high-irradiance lasers with matter is
described in more details in recent reviews
(Mourou {\it et al.\/} \cite{Mourou};
Salamin {\it et al.\/} \cite{bib:Salamin};
Marklund \& Shukla \cite{bib:Marklund}).

The dimensionless amplitude of the laser pulse at focus, $a_0$,
is related to the laser peak intensity by
$I_{0L} = a_0^2 \times 1.37\times 10^{18}$ W/cm$^2 \times (\mu{\rm m}/\lambda)^2$ 
for linear and $I_{0C} = 2 I_{0L}$ for circular polarization.
Other important dimensionless parameters characterizing the laser pulse
are the number of cycles corresponding to the pulse duration, $\tau_{\rm las}$,
as $N_{\rm las} = c \tau_{\rm las} / \lambda$, 
and
the focal spot diameter in terms of laser wavelengths, $D_{\rm las}/\lambda$.
Further, in the interaction of a laser pulse with plasma,
a collective response of plasma to a periodic EM wave
is characterized by the ratio
of the plasma electron number density, $n_e$, to critical density,
$n_{cr}=m_e\omega^2/4\pi e^2$, determined by the laser frequency.
It is related to the Langmuir frequency $\omega_{pe} = (4\pi n_e e^2/m_e)^{1/2}$
via $n_e/n_{cr} = \omega_{pe}^2/\omega^2$.
These parameters appear in the dispersion equation of
a relativistically strong EM wave with circular polarization
propagating in plasma,
$\omega^2 = k^2 c^2 + \omega_{pe}^2 (1+a_0^2)^{-1/2}$,
where $k$ is the wave number.


As seen in the experiment, theory and computer simulation, 
irradiation of various targets -- particle bunch, gas, liquid, porous target, solid -- 
by a relativistically strong laser pulse 
results in 
a formation of plasmas, laser pulse self-focusing, its frequency and shape modulation,
and a generation of strong collective EM fields 
leading to transformation of the laser pulse energy
into the energy of
(1) waves of different kinds, such as plasma wake wave and electromagnetic soliton;
(2) large-scale electrostatic field and quasi-static magnetic field;
(3) EM waves in different frequency range, such as terahertz radiation, 
      high harmonics up to ultraviolet (UV) and eXtreme ultraviolet (XUV), X-ray and $\gamma$-ray;
(4) plasma particles (electrons, positive and negative ions);
(5) products of nuclear reactions (e. g., neutrons, positrons, muons, etc.).
The observation of these new phenomena pointed the way to many promising applications.
The corresponding processes can be utilized for creation of useful tools in
laboratory setting (such as for diagnostics) and, much beyond specialized
scientific needs, in technology.

\begin{figure}[h]
\includegraphics[scale=0.75]{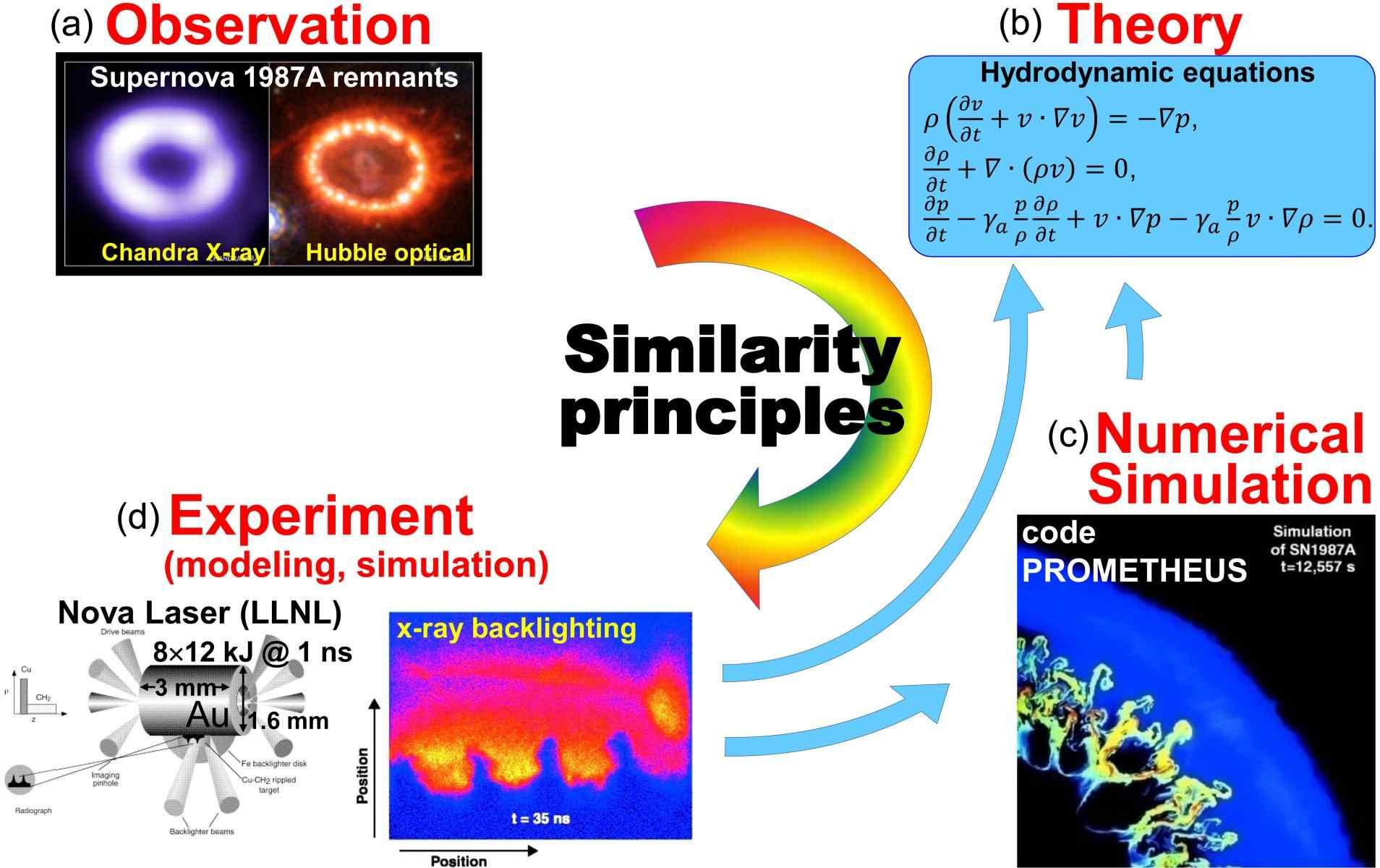}
\caption{\label{fig:SimilarityPrinciple}%
{\it Process simulation} with high-power laser-plasma interactions (d) and 
{\it configuration simulation} with computer models (c)
help to refine the theory (b) of observed phenomena (a),
as exemplified by the investigation of
the supernova shells instabilities.
Credit: 
(a) X-ray: NASA/CXC/PSU/S. Park \& D. Burrows, optical: NASA/STScI/CfA/P. Challis;
(c) Muller {\it et al.\/} \cite{bib:f11c};
(d) Remington {\it et al.\/} \cite{bib:f11d}.
}
\end{figure}

\begin{figure}[h]
\includegraphics[scale=0.75]{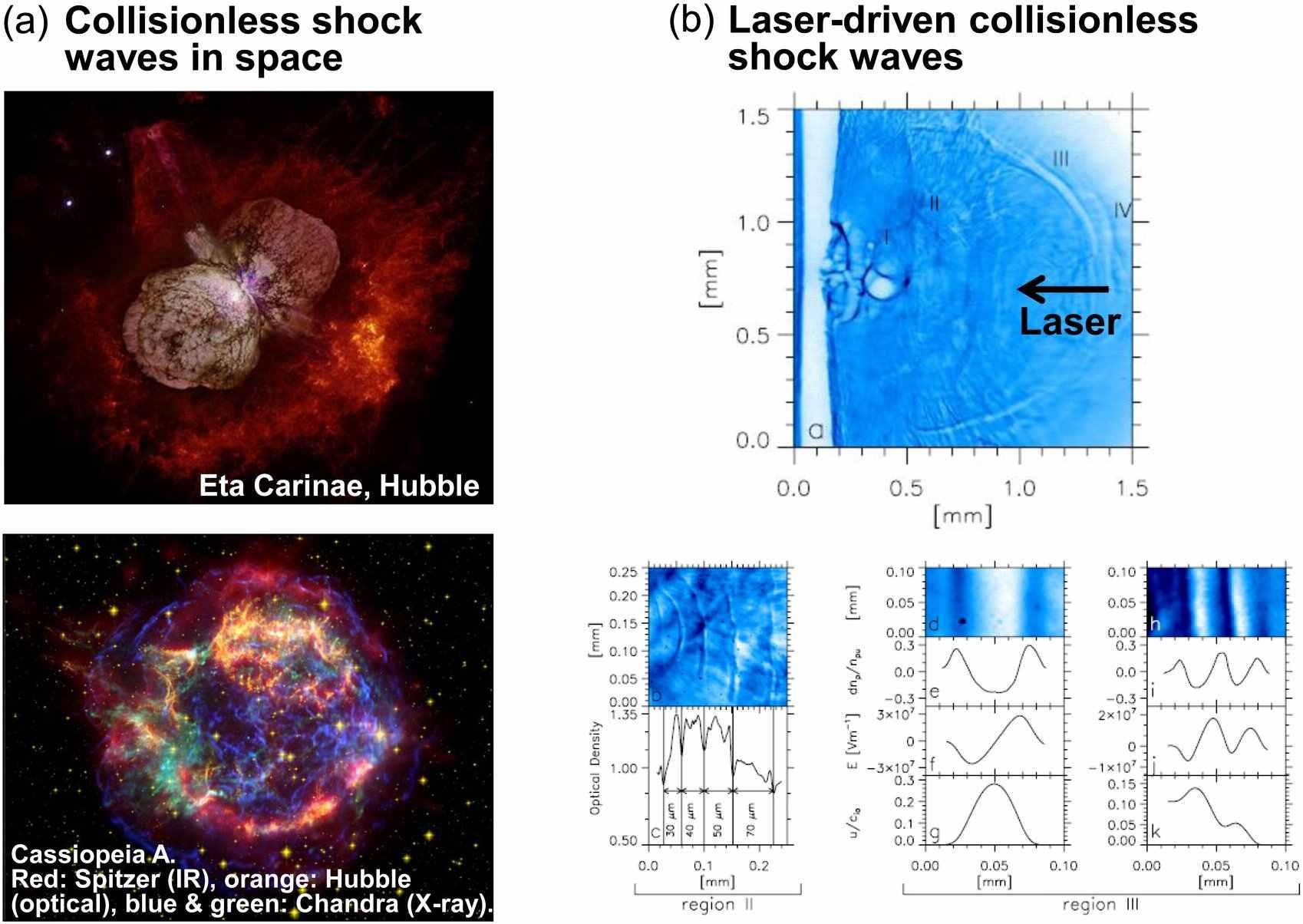}
\caption{\label{fig:S:CollShock}%
(a) Collisionless shock waves seen in space 
(credit: $\eta$ Carinae: NASA, ESA, N. Smith;
Cassiopeia A: X-ray: NASA/CXC/SAO, 
optical: NASA/STScI, infrared: NASA/JPL-Caltech/Steward/O. Krause {\it et al.\/}).
In laser plasma (b) the shock wave motion is detected with proton imaging
and studied with computer modeling (Romagnani {\it et al.\/} \cite{LShock}).
}
\end{figure}

\begin{figure}[h]
\includegraphics[scale=0.75]{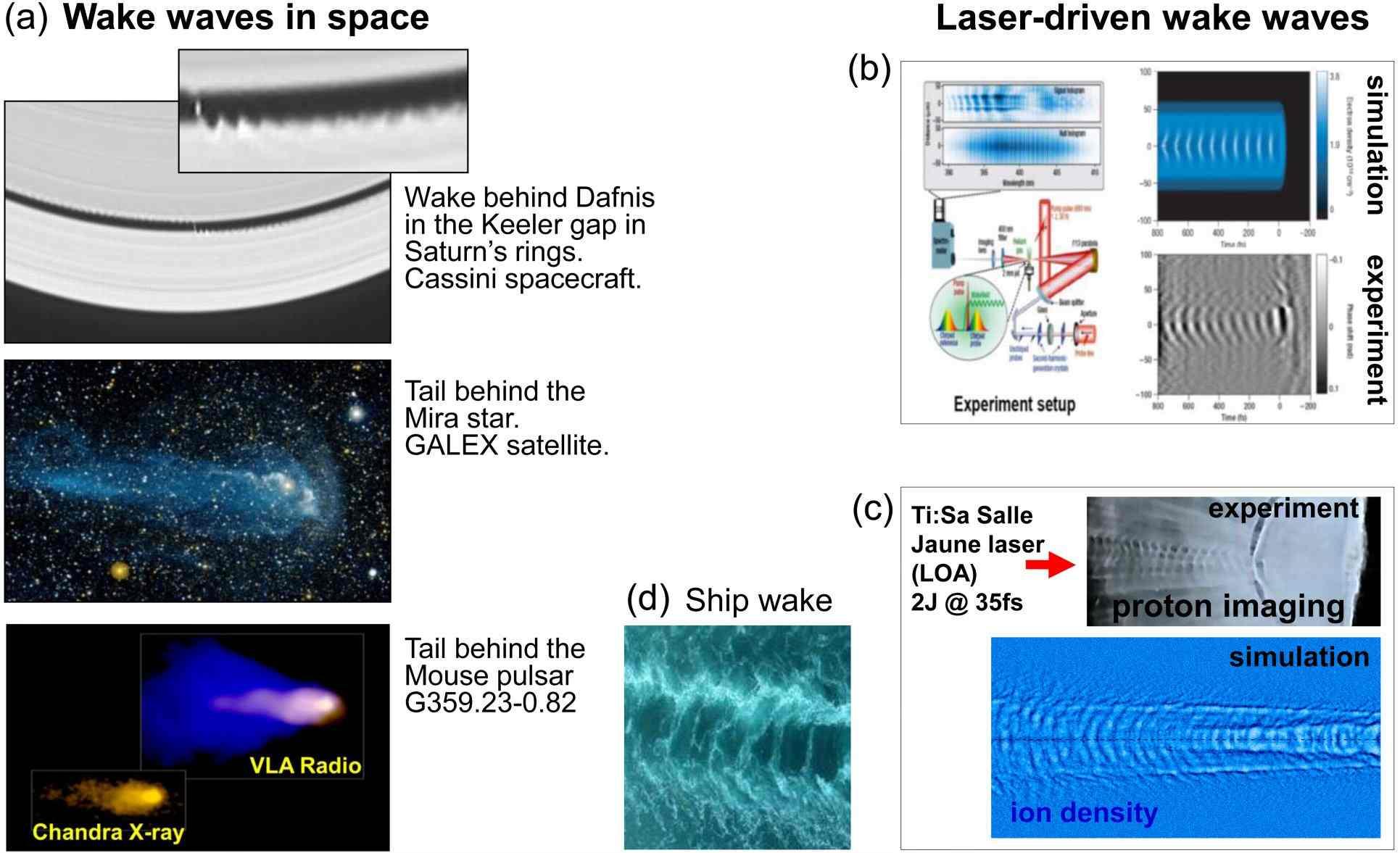}
\caption{\label{fig:S:Wake}%
Wake waves seen in space (a) and ocean (d).
In laser plasma wake wave manifests itself by a fast change of electron density (b)
snap-shot by frequency-domain holography
(Matlis {\it et al.\/} \cite{wakeelectron}) and
relatively slow response of ions (c)
imaged with MeV protons
(Borghesi {\it et al.\/} \cite{wakeion}).
(a) Credit: top: Cassini Imaging Team/SSI/JPL/ESA/NASA,
middle: NASA/JPL-Caltech/GALEX/C. Martin (Caltech) \& M. Seibert(OCIW),
bottom: NASA/CXC/SAO/B.Gaensler {\it et al.\/}, radio: NSF/NRAO/VLA.
}
\end{figure}

\begin{figure}[h]
\includegraphics[scale=0.75]{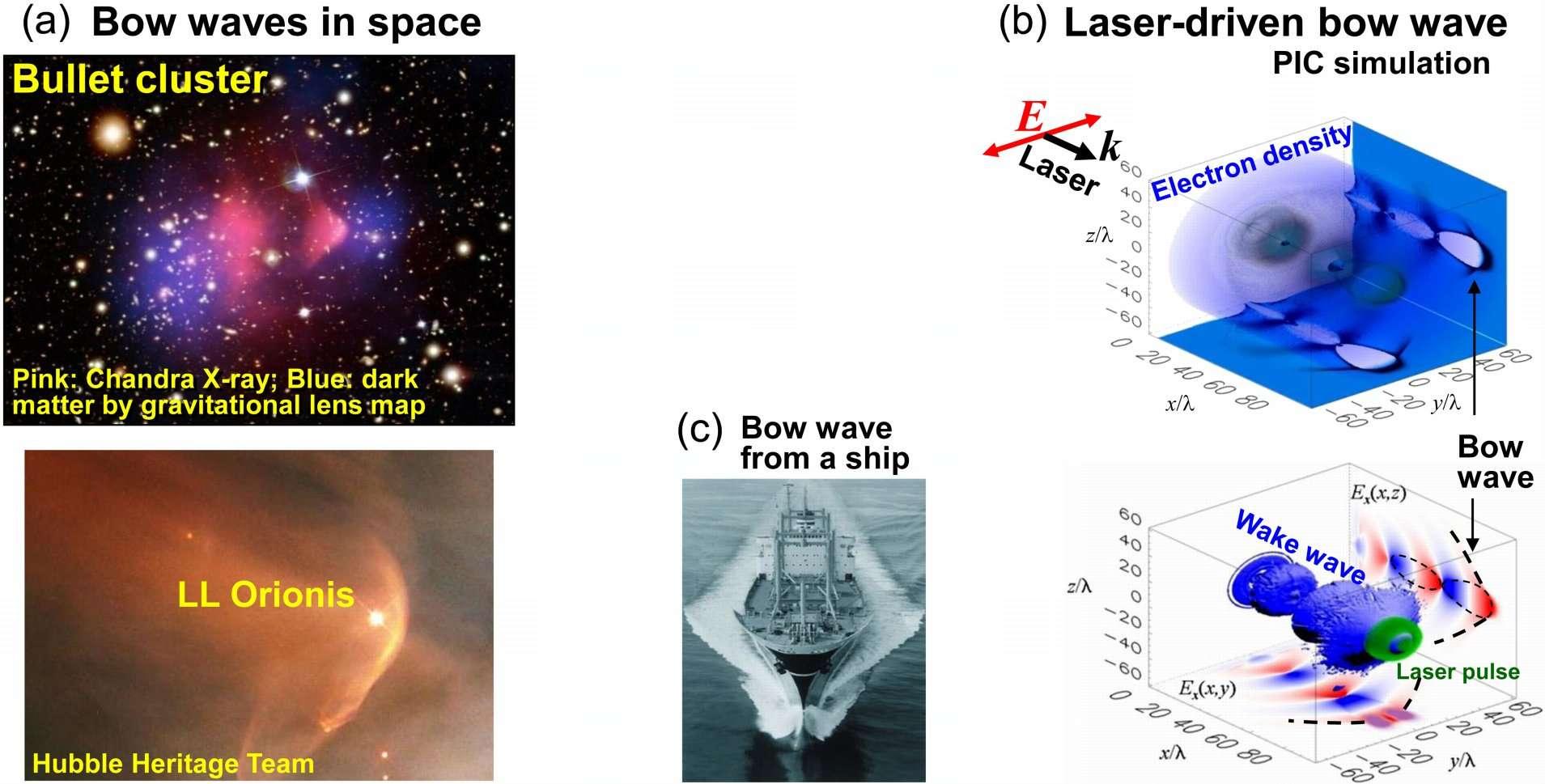}
\caption{\label{fig:S:Bow}%
Bow wave induced in ambient medium by a compact object as seen in space (a), 
ocean (c) and laser plasma simulations (b).
In plasma the role of the compact object is played by a tightly focused laser pulse
(Esirkepov {\it et al.\/} \cite{bow}).
(a) Credit: top: composite \& X-ray: NASA/CXC/CfA/M. Markevitch {\it et al.\/},
lensing map: NASA/STScI, ESO WFI, Magellan/U.Arizona/D. Clowe {\it et al.\/},
optical: NASA/STScI, Magellan/U.Arizona/D. Clowe {\it et al.\/};
bottom: Hubble Heritage Team (AURA/STScI), C. R. O'Dell (Vanderbilt), NASA.
}
\end{figure}

\section{Similitude}

Using a similarity of scaling laws between
the physical system and its model in a laboratory
and reproducing the key dimensionless parameters one can study
the corresponding physical process in a controllable way, Fig. \ref{fig:SimilarityPrinciple}.
Two systems behave identically
if they are described by the same equations 
with the same values of 
dimensionless parameters.
In astrophysical phenomena modeling, however,
such the identity can hardly be achieved
because the key dimensional parameters appear to be prohibitively large or small
and because an analytical theory of astrophysical phenomena is often far from being complete.
Fortunately, 
an approximate (incomplete) similarity with only few basic parameters 
is helpful for the development and refinement of the theory.
A so-called {\it configuration simulation} tries to reconstruct
the global geometry and some physical processes therein,
while a {\it process simulation} reproduces local properties of
physical processes at astrophysical conditions (F\"althammar \cite{bib:Falthammar}).
An important case of the approximate similarity
is {\it limited scaling} or {\it limited similarity},
where it is sufficient that 
``dimensionless quantities in nature which are small compared to
unity should be small in the model, but not necessarily by the same order
of magnitude'' (Block \cite{bib:Block}).

Laser-matter interactions provide many examples of
similarity with astrophysical phenomena and
allows an insight into fundamental processes at astrophysical conditions.

\begin{figure}[h]
\includegraphics[scale=0.75]{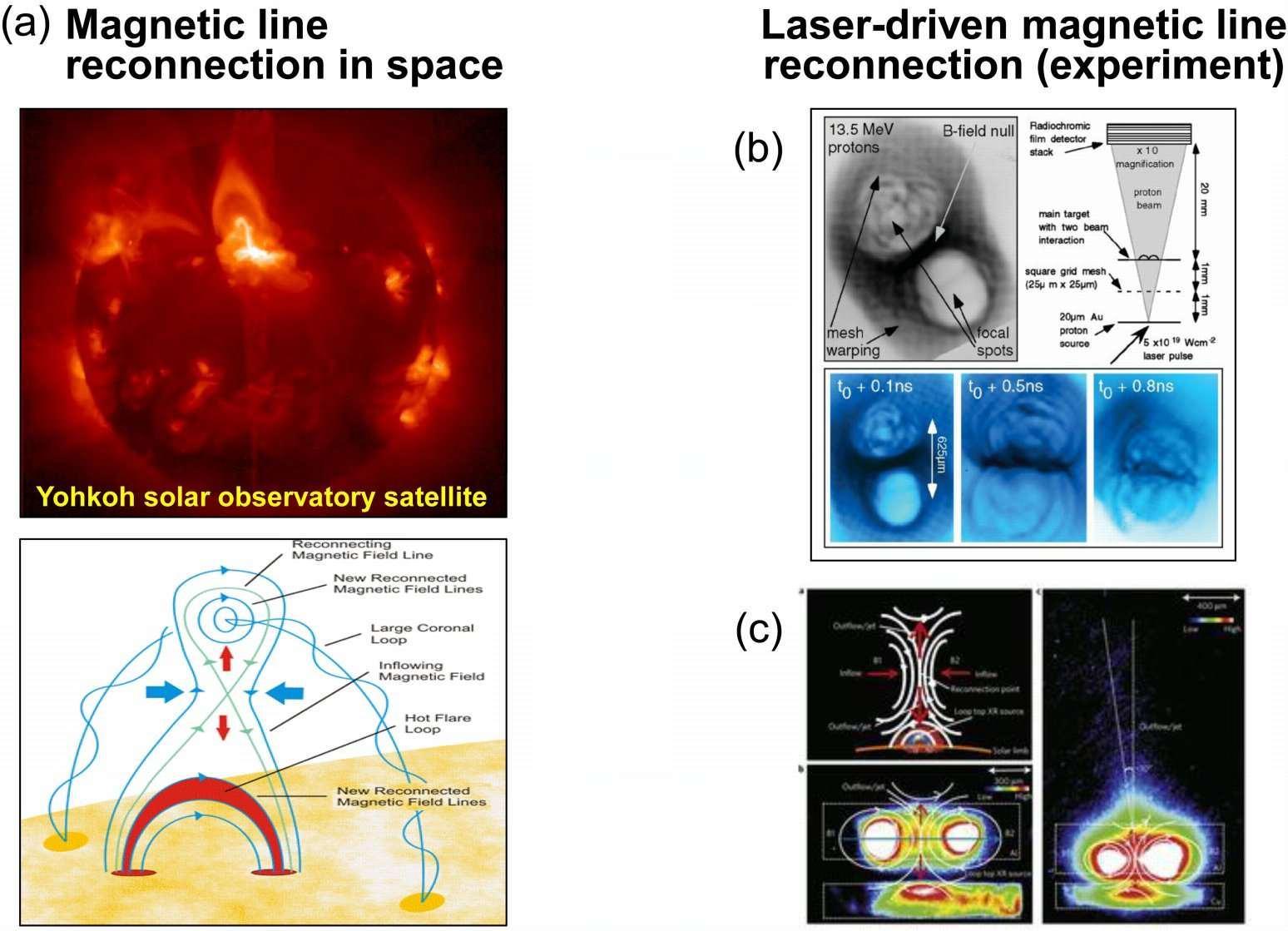}
\caption{\label{fig:S:Reconnect}%
The reconnection of magnetic field lines
seen (a) in solar flares 
(credit: ISAS, Yohkoh Project, SXT Group, K. Shibata {\it et al.\/})
and in high-power laser-plasma interactions
[(b) Nilson {\it et al.\/} \cite{MagRecExp1}; (c) Zhong {\it et al.\/} \cite{MagRecExp2}].
}
\end{figure}

\begin{figure}[h]
\includegraphics[scale=0.75]{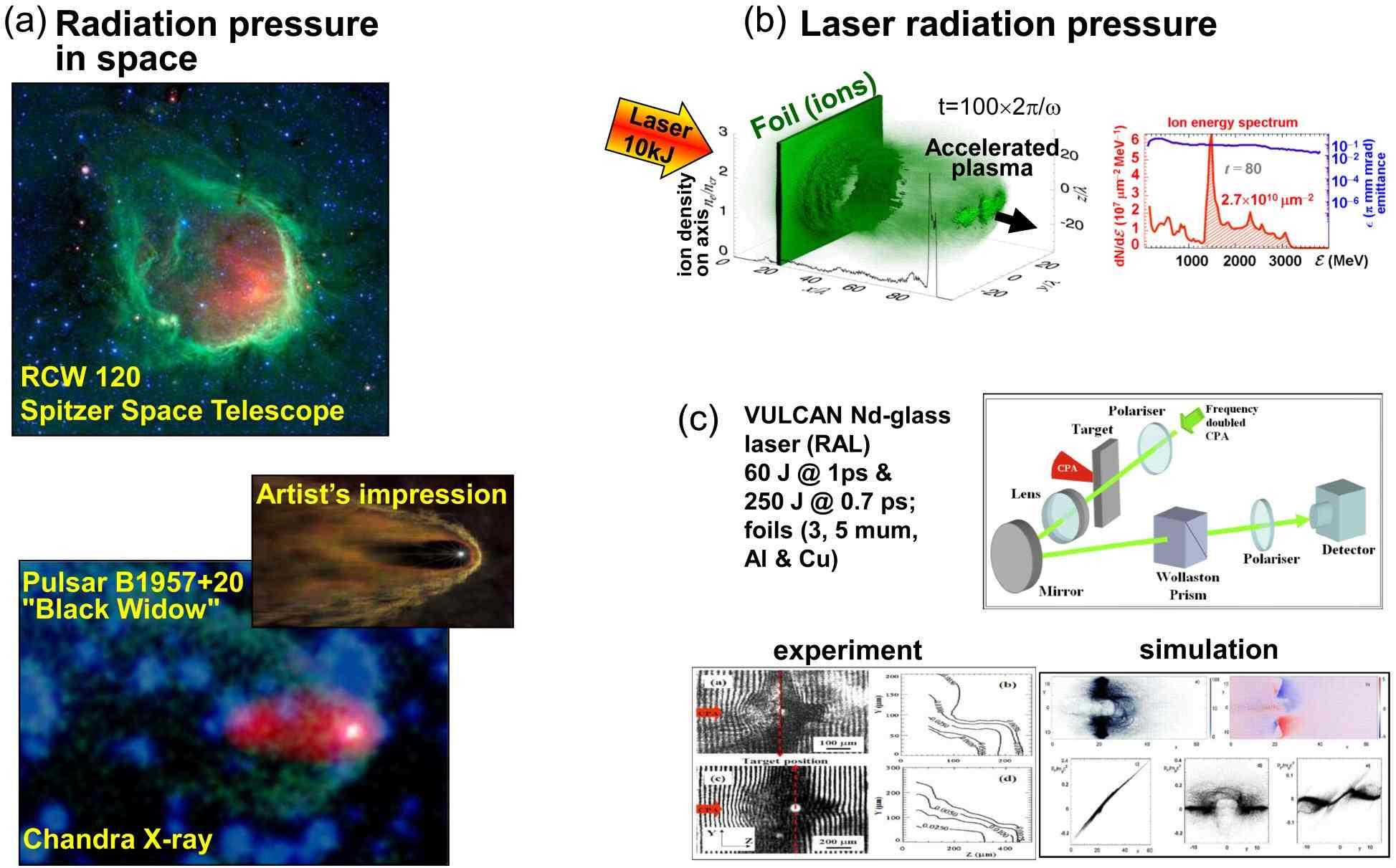}
\caption{\label{fig:S:RPDA}%
Electromagnetic radiation pressure revealed in space (a)
and in laser plasma with (b) computer simulation
(Bulanov SV {\it et al.\/} \cite{bib:RRD}; Esirkepov {\it et al.\/} \cite{RPDA})
and in (c) experiments (Kar {\it et al.\/} \cite{Jets}).
(a) Credit: top: NASA/JPL-Caltech/GLIMPSE-MIPSGAL Teams;
bottom: X-ray: NASA/CXC/ASTRON/B. Stappers {\it et al.\/}, 
optical: AAO/J. Bland-Hawthorn \& H.Jones,
artist's impression: NASA/CXC/M. Weiss.
}
\end{figure}

\begin{figure}[h]
\includegraphics[scale=0.75]{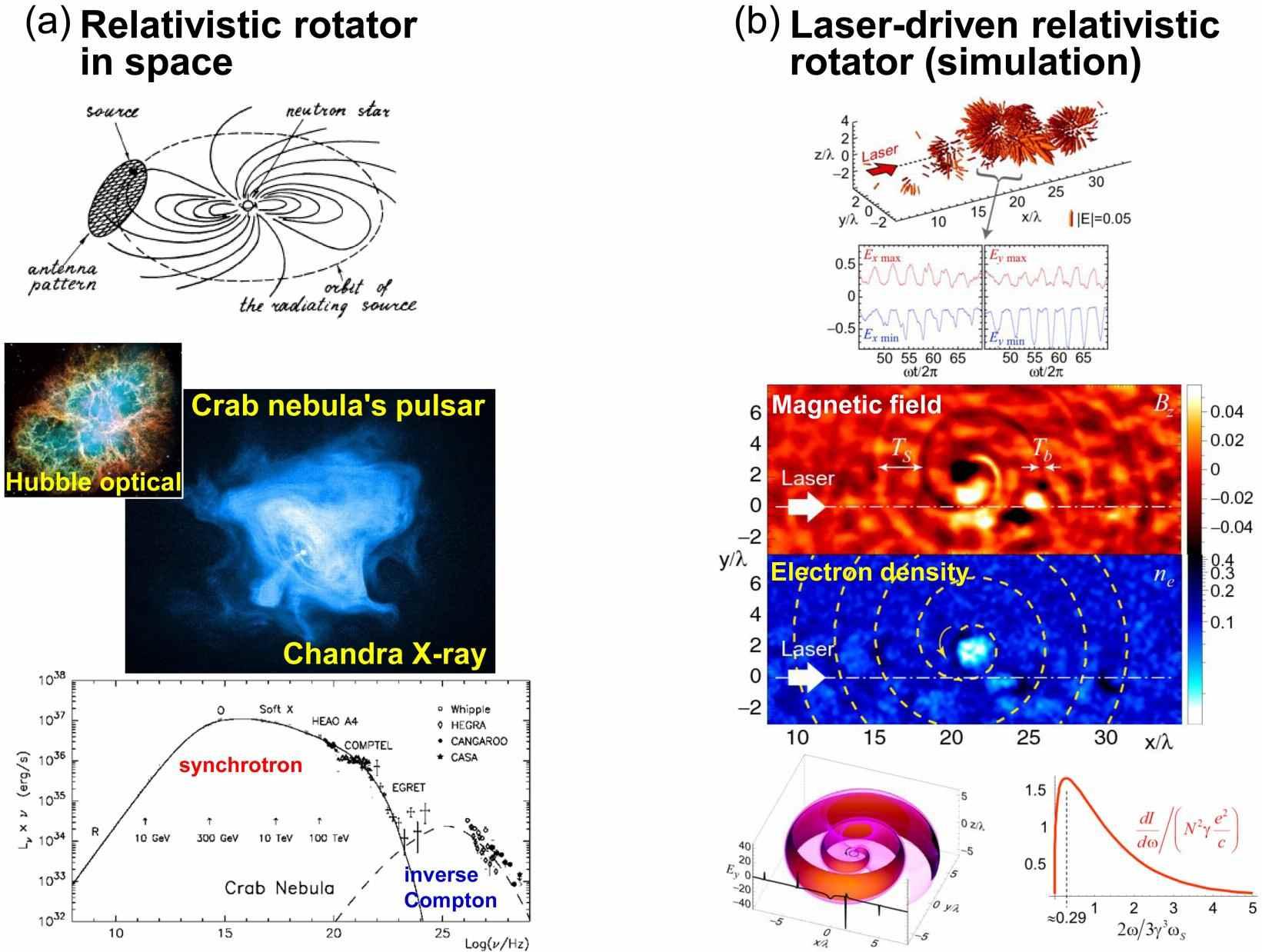}
\caption{\label{fig:S:Rotator}%
Relativistic rotators seen due to their emission
in space (a) and in laser plasma simulation (b),
where the emitter is a circularly polarized electromagnetic quasi-soliton
(Esirkepov {\it et al.\/} \cite{afterglow}).
(a) Credit: top: Ginzburg \& Zheleznyakov \cite{Ginzburg};
middle: optical: NASA/ESA/J. Hester \& A. Loll (ASU),
X-ray: NASA/CXC/SAO/F. D. Seward, W. H. Tucker \& R. A. Fesen;
bottom: Aharonian {\it et al.\/} \cite{crabSED}.
}
\end{figure}

Shocks developed during a supernova explosion
and Rayleigh-Taylor and Richt\-may\-er-Meshkov instabilities
of the supernova shells
create large scale modulations of the gas density.
The same type of instabilities are typically
developed in the interaction of high-power lasers with dense materials.
Laboratory experiments with laser-driven shocks
in high density plasma 
reveal 
fundamental properties of materials such as opacities and equations of state,
and allow
refining of numerical simulations and analytical models
for supernovae shocks development and evolution
(Remington {\it et al.\/} \cite{SNlab}; Fig. \ref{fig:SimilarityPrinciple}).
Underlying one of the mechanisms of cosmic ray acceleration,
shock waves propagate in interstellar plasma.
On later stages of evolution, when their amplitude is relatively small,
they can be described by Korteweg--de Veries--Burgers equation.
In high intensity laser-plasma experiments, Fig. \ref{fig:S:CollShock}, 
the collisionless shock structure and corresponding electric field 
have been studied with high spatial and temporal resolution
owing to proton imaging technique (Romagnani {\it et al.\/} \cite{LShock}).

Relative motion of compact objects and interstellar media
produce various wakes and (shock) bow waves.
In laboratory plasma these structures can be simulated to some extent
by a relativistically strong laser pulse creating
wake waves (Matlis {\it et al.\/} \cite{wakeelectron}; 
Borghesi {\it et al.\/} \cite{wakeion}; Fig. \ref{fig:S:Wake})
and bow waves (Esirkepov {\it et al.\/} \cite{bow}, Fig. \ref{fig:S:Bow})
in underdense plasma.

The magnetic field lines reconnection
visible in solar flares 
leads to a fast magnetic energy release 
into motion and heating of plasma, EM radiation emission
and charged particle acceleration.
The magnetic recconection phenomena, Fig. \ref{fig:S:Reconnect}, 
can be studied in the laboratory using high-power lasers 
(Nilson {\it et al.\/} \cite{MagRecExp1}; Zhong {\it et al.\/} \cite{MagRecExp2}).

Radiation pressure plays important role in astrophysical environments.
High intensity ultrashort laser pulses irradiating
thin foils can produce radiation pressure 
which dominates the plasma dynamics,
leading to efficient collective acceleration of ions 
to relativistic energy 
(Bulanov SV {\it et al.\/} \cite{bib:RRD};
Esirkepov {\it et al.\/} \cite{RPDA};
Kar {\it et al.\/} \cite{Jets}; Fig. \ref{fig:S:RPDA}).

Relativistically rotating compact objects
emitting periodic pulses of high-frequency radiation
are exemplifyed in astrophysics by pulsars.
In laser plasma a relativistic rotator can be modeled with
a circularly polarized electromagnetic quasi-soliton
(Esirkepov {\it et al.\/} \cite{afterglow}; Fig. \ref{fig:S:Rotator}).

\section{Reaching astrophysical conditions with extreme power lasers}

Extreme EM fields and ultra-high energy and matter densities
are present in various astrophysical objects,
as follows from the
observation of ultra-high-energy cosmic rays (Abraham {\it et al.\/} \cite{AugerCol}),
gamma ray bursts (Vedrenne \& Atteia \cite{GRB}),
radiation from pulsars (Lorimer \& Kramer \cite{Pulsar}), etc.
Extreme power lasers such as {\sf HiPER} and {\sf ELI}
enable 
the {\it process simulation} of matter dynamics 
at ultra-high energy density in astrophysical compact objects
and in the early universe.
With the laser-driven charged particle acceleration,
present-day lasers are capable of producing physical conditions 
which where accessible before in laboratory
only in high energy physics in the regime of a few particles interaction.
Combining multiple high-power laser beams into a tight focus
(Bulanov SS {\it et al.\/} \cite{multibeam})
or using relativistic mirrors formed by plasma wake waves
(Bulanov SV {\it et al.\/} \cite{FM}),
one can build up the critical field of quantum electrodynamics (Schwinger field)
in a ``macroscopic'' volume
(with the scale length much greater than the Compton wavelength of the electron).
This gives a tool for studying collective effects of quantum electrodynamics. 

\begin{figure}[h]
\includegraphics[scale=0.75]{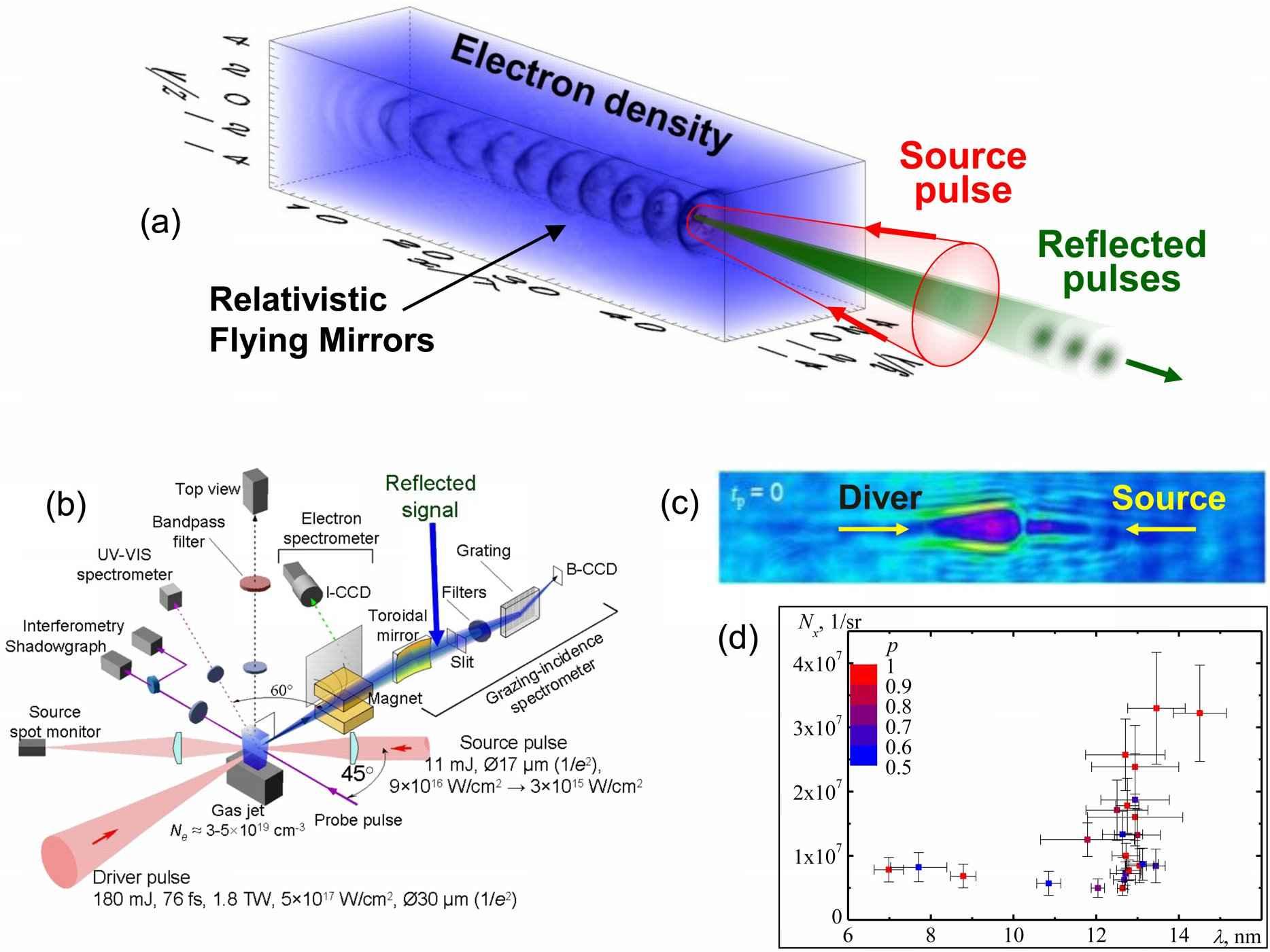}
\caption{\label{fig:RFM}%
The relativistic flying mirrors
formed by electron density modulations in the wake wave (a).
A counter-propagating laser beam is partially reflected
with a large frequency upshift.
The proof-of-priciple experiment (Kando {\it et al.\/} \cite{FMexp}):
setup (b),
colliding two pulses (c),
signal (d).}
\end{figure}

\subsection{Towards Schwinger field with relativistic flying mirror}

The change of EM radiation frequency 
emitted by or reflected from moving objects (Doppler effect)
is ubiquitous in astrophysical observations.
In laser plasma this effect can be used for the intensification
of EM radiation.

A relativistically strong laser pulse (driver) propagating in low-density plasma
efficiently creates a wake wave  (Akhiezer \& Polovin \cite{Akh-Pol}; Tajima \& Dawson \cite{TD79})
when the pulse duration is much less than the
inverse Langmuir frequency of plasma, $\omega_{pe}^{-1}$,
Fig. \ref{fig:RFM}.
The wake wave phase velocity $v_{ph}=\beta _{ph}c$ equals the laser pulse
group velocity which is close to the speed of light in vacuum.
The corresponding Lorentz factor is 
$\gamma_{ph}=(1-\beta_{ph}^{2})^{-1/2}\approx 
\omega/\omega_{pe}$, where $\omega$ is the laser pulse frequency.
When the laser dimensionless amplitude is sufficiently large,
$a_0 \ge (4n_{cr}/n_e)^{1/3}$ (Zhidkov \cite{Zh2004}),
the wake wave is breaking.
Near breaking the electron density profile in the wake takes the shape
of spikes separated by the wake wavelength, 
$\lambda_p \approx 4(2\gamma_{ph})^{1/2}c/\omega_{pe}$ for $\gamma_{ph}\gg 1$.
Due to the dependence of the wake wave frequency on the driver
pulse amplitude, in the transverse direction the spikes continue
as thin approximately paraboloidal shells (Bulanov SV \& Sakharov \cite{wake-parab}),
Fig. \ref{fig:RFM}.
These shells can partially reflect and focus
a counter-propagating laser pulse (source).
Due to a strong localization of the electron density in the shells,
for sufficiently long EM waves
the geometric optics approximation fails and
the reflection coefficient is not exponentially small.
The reflected wave vector-potential scales as $\gamma_{ph}^{-3/2}$
while the reflected energy flux grows as $\gamma_{ph}$
due to a large frequency upshift, $\omega_r \approx 4\gamma_{ph}^2 \omega$.
In addition, due to focusing the reflected wave intensity
is increased by the factor of
$8(D/\lambda)^2\gamma_{ph}^3$
(at focus) with respect to the incident wave,
where $D$ is the diameter of the reflecting shell.
This favorable scaling allows reaching Schwinger field
at which $e^-e^+$ pairs are created from vacuum
with present-day laser technology (Bulanov SV {\it et al.\/} \cite{FM}).
The frequency upshift of the laser pulse reflected
from relativistic flying mirrors
formed by the plasma wake wave 
has been observed in proof-of-principle experiments 
(Kando {\it et al.\/} \cite{FMexp};
Pirozhkov {\it et al.\/} \cite{FMexp-rev};
Kando {\it et al.\/} \cite{FMexp2}).

The curvature of the relativistic flying mirror is 
controlled by the transverse profile of the driver pulse.
The EM wave in the beam reflected from a flat flying mirror 
is almost planar and can not create $e^-e^+$ pairs,
since both the fundamental invariants of the EM field,
${\bf E}^2 - {\bf B}^2$ and ${\bf E}\cdot {\bf B}$, are zero.
In the case of ideally parabolic flying mirror,
the reflected beam focuses into a moving spot and then diverges
whithin the collimation angle of $\sim 1/\gamma_{ph}$.
The first fundamental invariant is of the order of ${\bf E}^2/\gamma_{ph}^2$,
so that the magnitude of the reflected beam
can exceed Schwinger field by the factor of $\gamma_{ph}$
without making vacuum unstable.
Varying an angle of the collision of two such beams
one can study different regimes of the $e^-e^+$ pairs creation from vacuum.

\subsection{Laser-driven electron-positron-gamma plasma}

Astrophysical observations of EM radiation
from compact objects such as 
pulsars and active galactic nuclei
reveal the important role of
the processes of 
$e^-e^+$ pairs creation from vacuum in strong EM fields
and abundant gamma-ray generation.
Understanding the mechanisms of vacuum breakdown and polarization is
challenging for nonlinear quantum field theories and for astrophysics
(Ruffini {\it et al.\/} \cite{Ruf2010}).
These fundamental processes
occuring in neutron star magnetospheres, 
sources of gamma ray bursts, 
and during the lepton epoch of the early universe,
can be produced in a terrestrial laboratory
with extreme power lasers using present day technology.

\begin{figure}[t]
\includegraphics[scale=0.5]{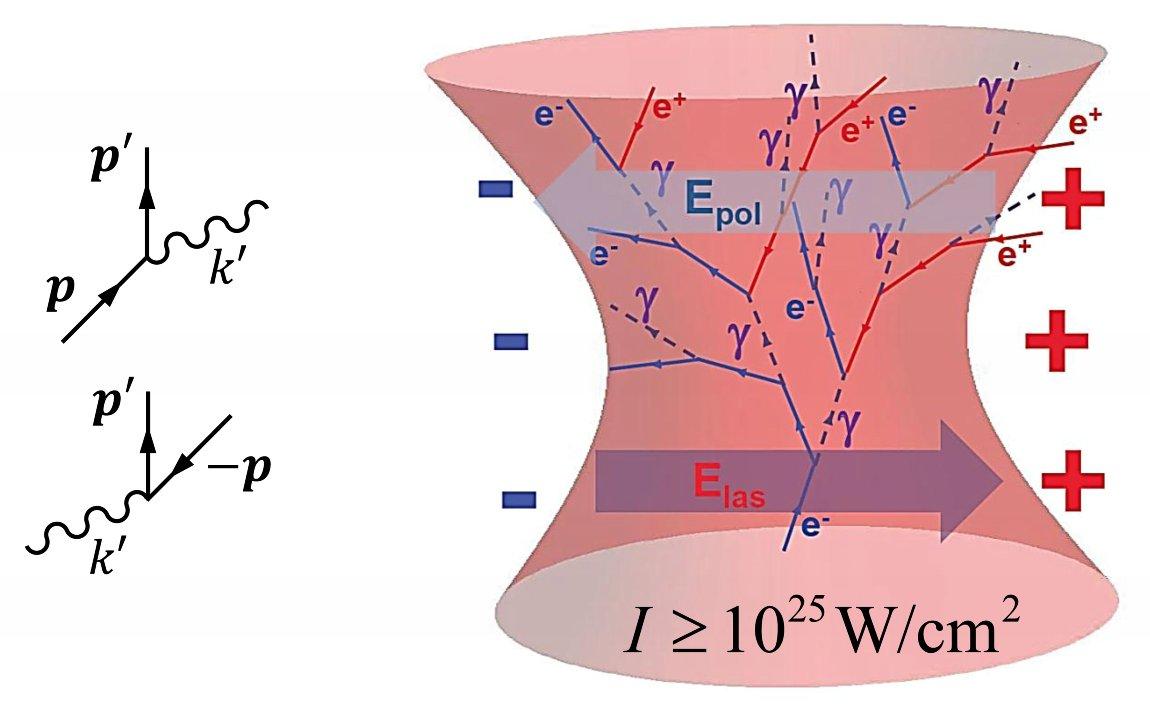}
\caption{\label{fig:EPGPlasma}%
Electron-positron-gamma plasma created at the focus
of extreme-power laser.}
\end{figure}

As mentioned in section \ref{sec:key},
the characteristic value of the electric field
in which vacuum becomes unstable
is $E_S = m_e^2c^3/e\hbar = 1.3\times 10^{16}$ V/cm
corresponding to the intensity of $I_S = 2.3\times 10^{29}$ W/cm$^2$.
However, at a laser focus with a few micron diameter
(much larger than the Compton wavelength of the electron)
the vacuum becomes unstable and
$e^-e^+$ pairs are created at two-three orders of magnitude less intensity,
well before the laser field reaches the Schwinger limit, 
due to a large phase volume occupied by a high-intensity EM field
(Bulanov SS {\it et al.\/} \cite{epbylaser}),
Fig. \ref{fig:EPGPlasma}.

If an electron (seed) get into a laser focus at the intensity
of the order of $10^{25}$ W/cm$^2$, 
it emits gamma photons
due to acceleration by the laser EM field
in the regime of a strong radiation friction.
Some gamma-photons emitted interact with photons comprising the laser field
and produce $e^-e^+$ pairs via multi-photon Breit-Wheeler process
(Breit \& Wheeler \cite{B-W}).
Then the pairs created generate another portion of gamma-photons,
powering an avalanche 
where {\it electron-positron-gamma plasma} ($e^-e^+\gamma$) is created
(Bell \& Kirk \cite{BellKirk};
Fedotov {\it et al.\/} \cite{Fedotov};
Bulanov SS {\it et al.\/} \cite{BulanovSS}).
The intensity threshold for the avalanche strongly depends on the
laser field polarization, being several orders of magnitude higher
in the case of linear polarization, which thus allow
an observation of $e^-e^+$ pair creation from vacuum
without seed electrons and avalanche (Bulanov SS {\it et al.\/} \cite{BulanovSS}).

In addition to the dimensionless amplitude of the electric field,
in QED new key parameters emerge in the description
of the charged particle interaction with EM fields.
The parameter
characterizing the probability of the photon emission by the electron
moving with 4-momentum $p$
in the the EM field given by the tensor $F^{\mu\nu}$,
corresponding to electric field strength $E$,
is
$\chi_e = e\hbar |F^{\mu\nu}p_\nu|/m_e^3c^4 \approx (E/E_S) (2p_\perp/m_e c) $.
Here $p_\perp$ is the electron momentum component perpendicular to the electric field.
In the electron rest frame of reference, $\chi_e \sim E/E_S$.
Another parameter
characterizes the probability of the $e^-e^+$ pair creation due to
the interaction of the EM field with
a high energy photon with 4-momentum $\hbar k$ and energy $\omega_\gamma$:
$\chi_\gamma = e\hbar^2 |F^{\mu\nu}k_\nu|/m_e^3c^4 \approx (E/E_S) (2\hbar\omega_\gamma/m_e c^2) $.
The probability of the avalanche and $e^-e^+\gamma$ plasma generation
via the multi-photon Breit-Wheeler process
is high when these two parameters are close to unity.

\begin{figure}[t]
\includegraphics[scale=0.75]{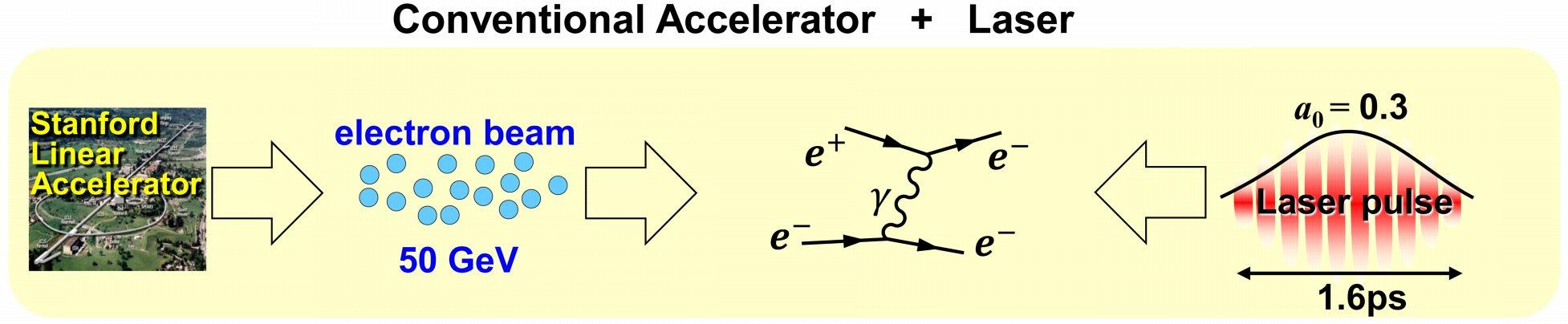}
\caption{\label{fig:SLAC}%
The experiment on the interaction of intense laser pulse
with electron bunch accelerated at the SLAC (Burke {\it et al.\/} \cite{SLAC96}).
Electron-positron pairs were created via
multiphoton inverse Compton scattering and multiphoton Breit-Wheeler process.}
\ \\
\includegraphics[scale=0.75]{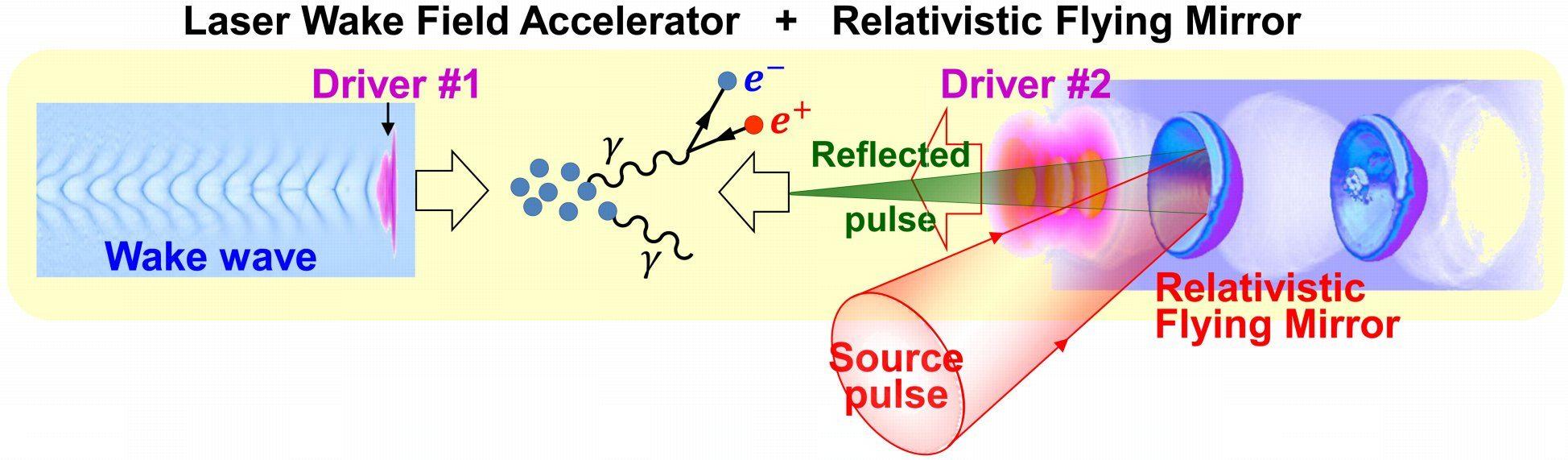}
\caption{\label{fig:LFWA-and-RFM}%
Proposed all-optical experiment on the electron-positron-gamma plasma creation
(Bulanov SV {\it et al.\/} \cite{Design}).
Laser wake field accelerator created by a laser pulse (driver 1) 
provides high-energy electron bunches that collides with an ultra-short
super-intense electromagnetic pulse from the relativistic flying mirror
created by two other laser pulses (driver 2 and source).}
\end{figure}

The requirements on the laser intensity securing $e^-e^+\gamma$ plasma creation
can be substantially relaxed in the interaction of the laser pulse with
a counter-propagating electron beam with sufficiently high energy.
In the experiments on the 527 nm terawatt laser interaction with 
46.6 GeV electrons from the SLAC accelerator beam,
positron generation was observed (Burke {\it et al.\/} \cite{SLAC96}),
Fig. \ref{fig:SLAC}. 
The laser photons were scattered by the electrons 
that generated high-energy photons
that collided with other laser photons to produce an electron-positron pair.
These two steps correspond to multiphoton inverse Compton scattering
and multiphoton Breit-Wheeler process, respectively.
In the reference frame of electrons the electric field
magnitude of the incident radiation was approximately 25\% of 
Schwinger field.
The values of the key parameters introduced above where
$\chi_e \approx 0.3$ and
$\chi_\gamma \approx 0.15$.

Recent advances in the developing of
the laser wake field accelerator (LWFA) of electrons
(Tajima \& Dawson \cite{TD79};
Leemans {\it et al.\/} \cite{Leemans};
Hafz {\it et al.\/} \cite{Hafz})
enabled an all-optical approach
(Bulanov SV {\it et al.\/} \cite{Design}),
Fig. \ref{fig:LFWA-and-RFM}.
A subpicosecond electron bunch with the energy of a few GeV
is accelerated in the wake wave generated by a petawatt laser pulse in plasma.
Electrons are injected from plasma
into the accelerating phase of the wake wave due to wave-breaking.
The accelerated electron bunch
collide with an extremely intense EM pulse generated by the
flying mirror, produced by two other laser pulses.
(In principle, the driver pulse for the LWFA can be the source for the flying mirror).
The setup can be optimally synchronized with femtosecond accuracy.
As in the case of the experiment at SLAC,
the collision between the electron bunch from the LWFA 
and the EM pulse from the flying mirror
produces gamma-photons via Thomson or inverse Compton scattering
and then the gamma-photons interact with the laser field
generating $e^-e^+$ pairs.
For 1.25 GeV electrons,
the source pulse amplitude, focal spot size and duration of
$a_{src}=1$, $D_{src}=10$ $\mu$m and $\tau_{src}=30$ fs, respectively,
and the flying mirror Lorentz factor of $\gamma_{FM}=5$,
we obtain that the EM field stregth at the focus of the flying mirror, $E_{FM}$,
increases with respect to the field strength in the source pulse, $E_{src}$,
as
$E_{FM} =2^{5/2}\gamma_{FM}^{3/2}(D_{src}/\lambda_0)E_{src} \approx 1.5\times 10^{-3} E_S$,
and the key parameters 
are
$\chi_e \approx 2\gamma_e E_{FM}/E_S = 7.5$ 
and 
$\chi_\gamma \approx (2\hbar \omega_\gamma/m_e c^2)(E_{FM}/E_S)
\approx 1.2\gamma_e E_{FM}/E_S = 4.5$.

\subsection{Laser-driven collider for quark-gluon plasma studies}

EM radiation pressure is important in astrophysical environments,
e.~g. it is dominating in formation of nebula surrounding $\eta$ Carinae
(van Boekel, R., {\it et al.\/} \cite{etaCarinae}).
Multi-petawatt femtosecond laser pulses
exert radiation pressure
which can dominate in the interaction with matter
(Bulanov SV {\it et al.\/} \cite{bib:RRD})
realizing one of the most efficient mechanisms
of ion acceleration
(Esirkepov {\it et al.\/} \cite{RPDA}).
This radiation pressure dominant acceleration (RPDA)
produces high quality ion bunches with relativistic energy
and high luminocity,
providing physical conditions 
for quark-gluon plasma generation
(Ludlam \& McLerran \cite{QGP}).

\begin{figure}[b]
\includegraphics[scale=0.75]{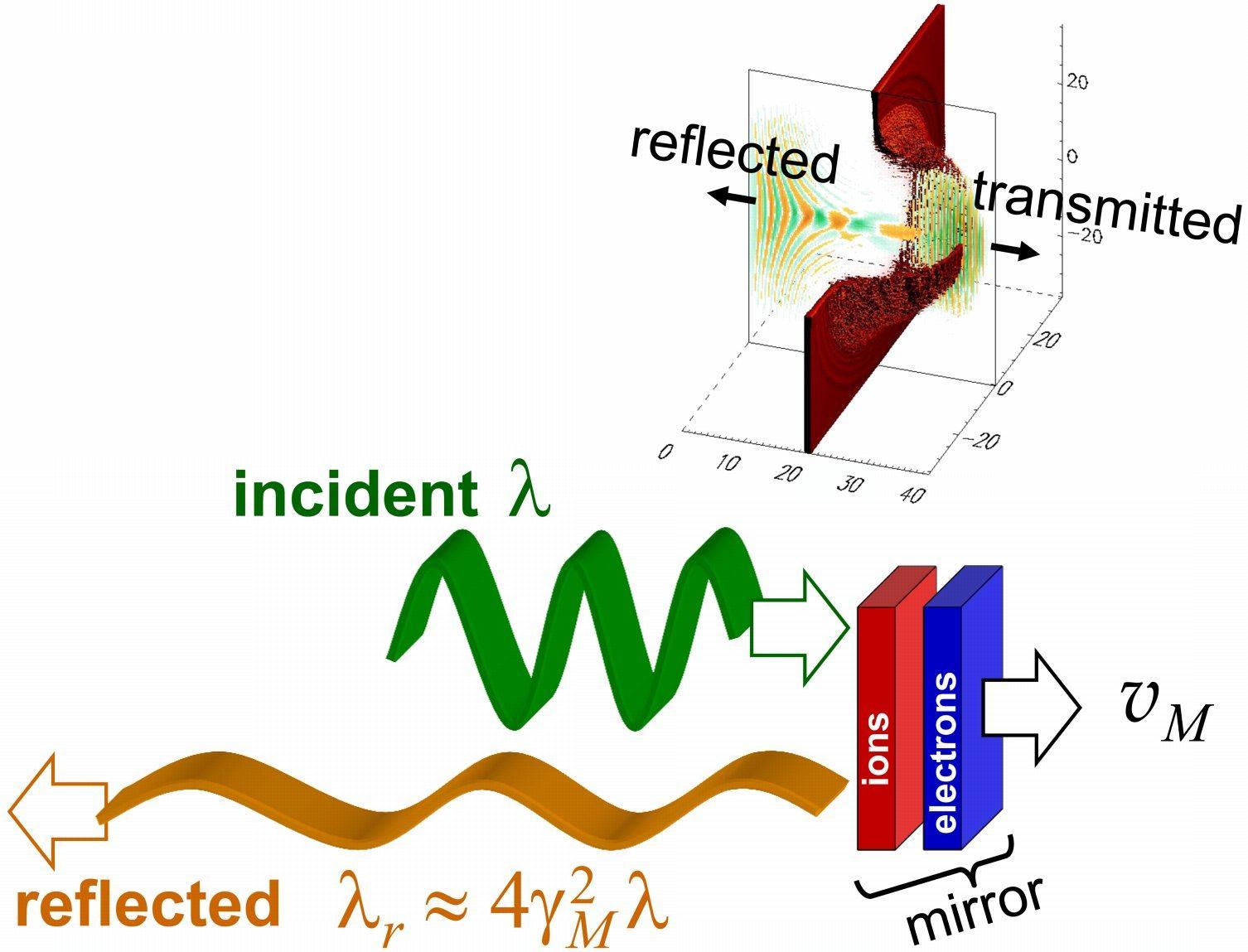}
\caption{\label{fig:RPDA}%
Radiation pressure dominant acceleration of ions.
On the top: 
the laser pulse is reflected with a gradual decrease of wavelength
while ions get accelerated as seen in numerical simulations
(Esirkepov {\it et al.\/} \cite{RPDA}).}
\end{figure}

\begin{figure}[t]
\includegraphics[scale=0.75]{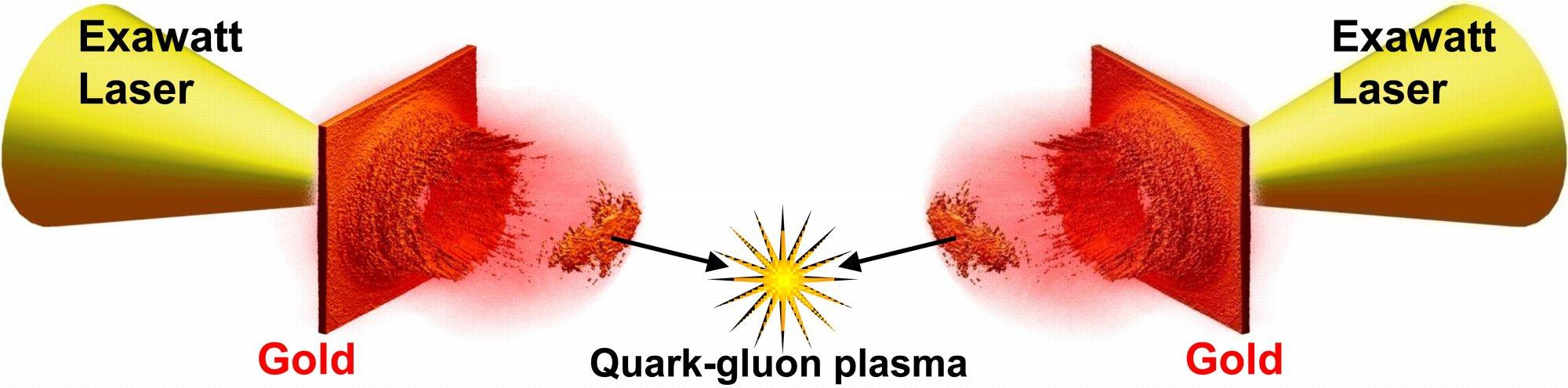}
\caption{\label{fig:Collider}%
Laser-driven collider for quark-gluon plasma studies.}
\end{figure}

The RPDA is realized when an extreme power laser pulse
pushes forward an irradiated region of a thin foil.
Plasma is accelerated as a whole,
but ions aquire almost all the energy
because of a large ion-to-electron mass ratio.
The accelerated foil soon become relativistic
and acts as a receding mirror,
reflecting the laser pulse, Fig. \ref{fig:RPDA}.
Since the reflected wave frequency decreases by the factor
of $4\gamma_M^2$, where $\gamma_M$ is the mirror Lorentz factor,
the laser energy is almost completely transformed into the energy of fast ions. 
The accelerated foil motion is described by the equation
$d/dt\{p_i + 2[p_i^3+(m_i^2c^2+p_i^2)^{3/2}]/3m_i^2c^2\} = E_0^2/2\pi n_e l $,
where $p_i$ and $m_i$ are the ion momentum and mass, 
$n_e$ and $l$ are the electron density and thickness of the foil, respectively.
By analogy with the self-consistent motion of the electron undergoing Thomson scattering
in an EM wave (Landau \& Lifshitz \cite{LL}),
while in our case the role of the Thomson cross section is played by the quantity $2/n_e l$,
the asymptotic energy gain is
${\cal E}_i \approx m_i c^2 (e E_0^2 t /8\pi n_e l m_i c )^{1/3}$.
The laser-to-ion energy transfromation efficiency is
${\cal N}_i {\cal E}_i / {\cal E}_L = 2{\cal E}_L/(2{\cal E}_L + {\cal N}_i m_i c^2)$,
where ${\cal E}_L$ is the laser pulse energy and ${\cal N}_i$ is the ion number.
If the laser pulse is sufficiently long, almost all its energy is transformed into ions.

A collision of two heavy ion beams 
accelerated up to the energy over 100 GeV per nucleon
by exawatt laser pulses, Fig. \ref{fig:Collider},
can provide the same number of
events of quark-gluon plasma generation
in one shot
as the Relativistic Heavy Ion Coillider ({\sf \cite{RHIC}})
in few months (Esirkepov {\it et al.\/} \cite{RPDA}).

\subsection{Laser-driven beam dump facility for neutrino oscillations studies}

\begin{figure}[h]
\includegraphics[scale=0.75]{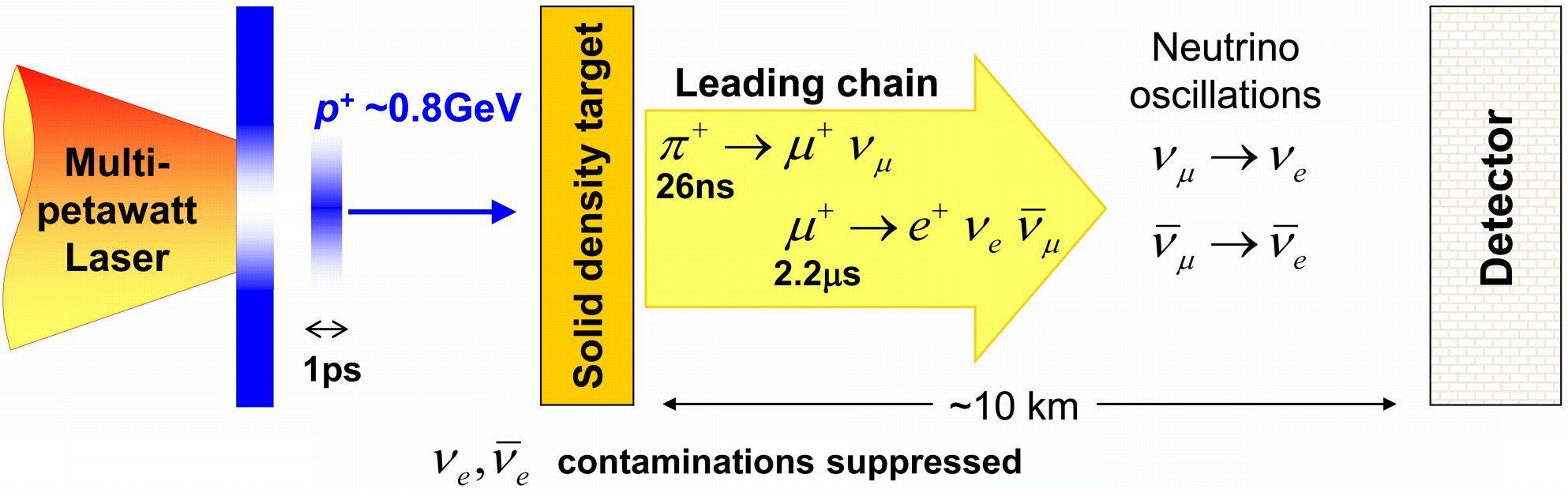}
\caption{\label{fig:Neutrino}%
Laser-driven beam dump facility for neutrino oscillations studies.}
\end{figure}

The radiation pressure dominant acceleration of protons in the GeV energy range
can give a source of low-energy (10-200 MeV) neutrinos,
which is suitable for neutrino oscillation studies
at the ``atmospheric'' scale (Bulanov SV {\it et al.\/} \cite{neutrino}).
In such the source
a high-current beam of protons is stopped by a solid density target,
producing pions, Fig. \ref{fig:Neutrino},
in analogy with 
the {\sf LAMPF} (presently {\sf \cite{LANSCE}}) 
beam dump facility supplied neutrinos for the {\sf LSND} experiment 
(Athanassopoulos {\it et al.\/} \cite{LSND}).
Neutrinos are created with energy of the order of a few tens of MeV
through $\pi^+\rightarrow\mu^+\nu_\mu$ decay-at-rest (DAR) chain 
and
through the subsequent $\mu^+\rightarrow e^+\nu_e\tilde\nu_\mu$ decay,
while the intrinsic $\nu_e$ ($\tilde\nu_e$) beam contamination 
can be kept below 0.1\% by keeping the proton energy
below the kaon and heavier mesons production threshold ($< 3$ GeV).
The oscillations
$\nu_\mu \rightarrow \nu_e$ and
$\tilde\nu_\mu \rightarrow \tilde\nu_e$
can be simultaneously observed at baselines of $L\sim 10$ km
(Bulanov SV {\it et al.\/} \cite{neutrino}).
Measuring the probabilities of these oscillations
at the ``atmospheric'' scale
one can refine the constrain on the
angle $\theta_{13}$ appearing in the
Pontecorvo-Maki-Nakagawa-Sakata (PMNS) matrix,
which characterizes mixing between different types of neutrino.
Such measurement is the most challenging task for neutrino oscillation experiments.
If $\theta_{13}$ is significantly smaller than the current limit ($\lesssim 10^o$),
traditional accelerating techniques will be unable to provide the required intensity
and purity. 
The laser-driven facility could allow the study of subdominant
$\nu_\mu \rightarrow \nu_e$ oscillations at the ``atmospheric'' scale for the first time.
In addition, such a facility can be developed in conjunction with projects 
for inertial confined nuclear fusion and neutron spallation sources
(Terranova {\it et al.\/} \cite{Nuclappl}).

\section{Conclusion}

Extreme-power lasers such as {\sf ELI} and {\sf HiPER} will allow 
{\it process simulation} for relativistic astrophysics
and will lead to new discoveries of collective phenomena in physics.
The state of matter peculiar to
cosmic gamma ray bursts,
the lepton epoch and hadron epoch of the early universe,
and processes, 
which were accessible on Earth only in the 
high energy physics experiments 
with conventional accelerators of charged particles,
can be produced with the help of 
super-intense lasers.
Thus the lasers entering the realm of extreme intensities 
become tools for driving and studying processes in the
{\it cosmology} domain of laboratory astrophysics.


\end{document}